\documentclass{article}
\usepackage{spconf,amsmath,graphicx}
\usepackage{url}
\usepackage{threeparttable}
\usepackage{multirow}
\usepackage{multicol}
\usepackage{booktabs}

\newcommand{\tabincell}[2]{\begin{tabular}
		{@{}#1@{}}#2\end{tabular}}

\title{THUEE system description for NIST 2019 SRE CTS Challenge}
\twoauthors
{\tabincell{c}{Yi Liu, Tianyu Liang, Can Xu, Xianwei Zhang,\\
		Xianhong Chen, Wei-Qiang Zhang, Liang He \sthanks{Co First Author,heliang@mail.tsinghua.edu.cn}}}
{Department of Electronic Engineering\\
	Tsinghua University, Beijing, China}
{\tabincell{c}{Dandan song, Ruyun Li, Yangcheng Wu,\\
		Peng Ouyang, Shouyi Yin}}
{Institute of Microelectronics,	\\
	Tsinghua University, Beijing, China\\
	and TsingMicro Co. Ltd.}

\begin{document}
	\ninept
	\maketitle
	\begin{abstract}
		This paper describes the systems submitted by the department of electronic engineering, institute of microelectronics of Tsinghua university and TsingMicro Co. Ltd. (THUEE) to the NIST 2019 speaker recognition evaluation CTS challenge. Six subsystems, including etdnn/ams, ftdnn/as, eftdnn/ams, resnet, multitask and c-vector are developed in this evaluation.
	\end{abstract}

	\section{Introduction}
	\label{sec:intro}
	This paper describes the systems developed by the department of electronic engineering, institute of microelectronics of Tsinghua university and TsingMicro Co. Ltd. (THUEE) for the NIST 2019 speaker recognition evaluation (SRE) CTS challenge \cite{sre19}. Six subsystems, including etdnn/ams, ftdnn/as, eftdnn/ams, resnet, multitask and c-vector are developed in this evaluation. All the subsystems consists of a deep neural network followed by dimension deduction, score normalization and calibration. For each system, we begin with a summary of the data usage, followed by a description of the system setup along with their hyperparameters. Finally, we report experimental results obtained by each subsystem and fusion system on the SRE18 development and SRE18 evaluation datasets.
	
	\section{Data Usage}
	For the sake of clarity, the datasets notations are defined as in table 1 and the training data for the six subsystems are list in table 2, 3, and 4.
	
	\begin{table}[!h]
		\begin{center}
			\caption{Datasets Notations}
			\begin{threeparttable}
				\begin{tabular}{|c|c|}
					\hline
					notation                & datasets                   \\
					\hline
					SRE                     & SRE04/05/06/08/10/MIXER6   \\
					\hline
					\multirow{2}{*}{SWB} 	& LDC98S75/LDC99S79/LDC2002S06/	 \\
					                        & LDC2001S13/LDC2004S07          \\
					\hline
					Voxceleb                &Voxceleb 1/2              \\
					\hline
					Fisher+SWB I            &Fisher + Switchboard I    \\
					\hline
					CH+CF                   &Callhome+Callfriend       \\
					\hline
				\end{tabular}
			\end{threeparttable}
		\end{center}
	\end{table}

	\begin{table}[!h]
		\begin{center}
			\caption{Data usage for etdnn/ams, ftdnn/as, and resnet subsystems}
			\begin{threeparttable}
				\begin{tabular}{|c|c|}
					\hline
					Components       & Data usage     \\
					\hline
					Neural Network   & SRE+SWB+Voxceleb \\
					\hline
					LDA/PLDA         & SRE+SRE16+SRE18 \\
					\hline
					PLDA-adapt       & SRE+SRE16+SRE18 \\
					\hline
					asnorm           & SRE18 unlabel \\
					\hline
				\end{tabular}
			\end{threeparttable}
		\end{center}
	\end{table}
	
	\begin{table}[!h]
		\begin{center}
			\caption{Data usage for multitask and c-vector subsystems}
			\begin{threeparttable}
				\begin{tabular}{|c|c|}
					\hline
					Components      & Data usage     \\
					\hline
					GMM-HMM         & Fisher+SWB I  \\
					\hline
					Neural Network 	& SRE+SWB+Voxceleb+Fisher+SWB I \\
					\hline
					LDA/PLDA        & SRE+SRE16+SRE18               \\
					\hline
					PLDA-adapt 	    & SRE16+SRE18+MIXER6+CH+CF \\
					\hline
					asnorm          & SRE18 unlabel \\
					\hline
				\end{tabular}
			\end{threeparttable}
		\end{center}
	\end{table}

	\begin{table}[!h]
		\begin{center}
			\begin{threeparttable}
				\caption{Data usage for eftdnn subsystem}
				\begin{tabular}{|c|c|}
					\hline
					Components      & Data usage     \\
					\hline
					Neural Network 	& SRE+SWB+Voxceleb+CH+CF \\
					\hline
					LDA/PLDA        & SRE+SRE16+SRE18 eval             \\
					\hline
					PLDA-adapt      & SRE+SRE16+SRE18 eval  \\
					\hline
					asnorm          & SRE18 unlabel \\
					\hline
				\end{tabular}
			\end{threeparttable}
		\end{center}
	\end{table}
	
	\section{Systems}
	
	\subsection{Etdnn/ams}
	Etdnn/ams system is an extended version of tdnn with the additive margin softmax loss \cite{AMsoftmax19}. Etdnn is used in speaker verification in \cite{JHUMITsre18}.  Compared with the traditional tdnn in \cite{Snyder2017Deep}, it has wider context and interleaving dense layers between each two tdnn layers. The architecture of our etdnn network is shown in table \ref{tab:etdnn}. It is the same as the etdnn architecture in \cite{JHUMITsre18}, except that the context of layer 5 of our system is t-3:t+3 instead of t-3, t, t+3. The x-vector is extracted from layer 12 prior to the ReLU non-linearity. For the loss, we use additive margin softmax with $m=0.15$ instead of traditional softmax loss or angular softmax loss. Additive margin softmax is proposed in \cite{AMsoftmax18} and then used in speaker verification in our paper \cite{AMsoftmax19}. It is easier to train and generally performs better than angular softmax.
	
	\begin{table}[!h]
		\begin{center}
			\begin{threeparttable}
				\caption{Etdnn architecture}
				\label{tab:etdnn}
				\begin{tabular}{ c|c|c|c}
					\hline
					Layer                & Layer Type     &Context &Size \\
					\hline
					1 	&  TDNN-ReLU    &t-2:t+2     &512 \\
					\hline
					2 	&  Dense-ReLU    &t              &512 \\
					\hline
					3 	&  TDNN-ReLU    &t-2,t,t+2     &512 \\
					\hline
					4 	&  Dense-ReLU    &t              &512 \\
					\hline
					5 	&  TDNN-ReLU    &t-3:t+3     &512 \\
					\hline
					6 	&  Dense-ReLU    &t              &512 \\
					\hline
					7 	&  TDNN-ReLU    &t-4,t,t+4    &512 \\
					\hline
					8 	&  Dense-ReLU    &t              &512 \\
					\hline
					9 	&  Dense-ReLU    &t              &512 \\
					\hline
					10 	&  Dense-ReLU    &t              &1500 \\
					\hline
					11 	&  Pooling(mean+stddev)    &Full-seq             &2$\times$1500 \\
					\hline
					12       &Dense(Embedding)-ReLU    &     &512. \\
					\hline
					13     &Dense-ReLU     &              &512. \\
					\hline
					14      &Dense-Softmax       &      & Num. spks. \\
					\hline
				\end{tabular}
			\end{threeparttable}
		\end{center}
	\end{table}

	\subsection{ftdnn/as}
	
	Factorized TDNN (ftdnn) architecture is listed in table \ref{tab:ftdnn}. It is the same to \cite{JHUMITsre18} except that we use 1024 nodes instead of 512 nodes in layer 12 and 13. The x-vector is extracted from layer 12 prior to the ReLU non-linearity. So our x-vector is 1024 dimensional. More details about the architecture can be found in \cite{JHUMITsre18}.
	\begin{table}[!h]
		\begin{center}
			\begin{threeparttable}
				\caption{ftdnn architecture}
				\label{tab:ftdnn}
				\begin{tabular}{ c|c|c|c|c|c|c}
					\hline
					\multirow{2}{*}{  }               &Layer      &Context   & Context          & conn.  & \multirow{2}{*}{Size} &Inner  \\
					& Type                                                    &factor 1    &factor 2    &from  &                                    &size  \\
					\hline
					1 	&  TDNN    &t-2:t+2          &        &       &512    &   \\
					\hline
					2 	&  F-TDNN    &t-2,t         &t, t+2    &      &1024   &256 \\
					\hline
					3 	&  F-TDNN   &t         &t   &       &1024   &256 \\
					\hline
					4 	&  F-TDNN   &t-3, t        &t, t+3   &      &1024   &256 \\
					\hline
					5 	&  F-TDNN    &t             &t         & 3     &1024   &256 \\
					\hline
					6 	&  F-TDNN    &t-3, t        &t, t+3   &      &1024   &256 \\
					\hline
					7 	&  F-TDNN   &t-3, t        &t, t+3   & 2,4     &1024   &256 \\
					\hline
					8 	&  F-TDNN    &t-3, t        &t, t+3   &      &1024   &256 \\
					\hline
					9 	&  F-TDNN    &t-3, t        &t, t+3   & 4,6,8     &1024   &256 \\
					\hline
					10 	&  Dense    &t  &t     &            &2048     & \\
					\hline
					11 	&  Pooling   &full-seq         & &    & 4096 & \\
					\hline
					12       &Dense    &  & &    &1024& \\
					\hline
					13     &Dense    &    & &          &1024 &\\
					\hline
					\multirow{2}{*}{14 }      &Dense-        &    &  &  & N. spks.  &\\
					&Softmax      &    &  &  &      &   \\
					\hline
				\end{tabular}
			\end{threeparttable}
		\end{center}
	\end{table}
	
	\subsection{eftdnn/ams}
	Extended ftdnn (eftdnn) is a combination of etdnn and ftdnn.
	Its architecture is listed in table \ref{tab:eftdnn}.
	The x-vector is extracted from layer 22 prior to the ReLU non-linearity.
	\begin{table*}[!h]
		\begin{center}
			\begin{threeparttable}
				\caption{eftdnn architecture}
				\label{tab:eftdnn}
				\begin{tabular}{ c|c|c|c|c|c|c|c}
					\hline
					\multirow{2}{*}{  }               &  Layer        &Context   & Context     & Context        & conn.    & \multirow{2}{*}{Size} &Inner  \\
					&Type     &factor 1    &factor 2       &factor 3     &from       &                                    &size  \\
					\hline
					1 	&  TDNN    &t-2:t+2          &        &   &     &512    &   \\
					\hline
					2 	&  Dense    &           &        &   &           &512    &   \\
					\hline
					3 	&  F-TDNN    &t-3,t -1        &t-1, t+1    &t+1, t+3 &      &1024   &256 \\
					\hline
					4         & Dense      &      &       &       &      &1024     & \\
					\hline
					5 	&  F-TDNN   &t         &t   &t  &       &1024   &256 \\
					\hline
					6       &Dense     & & & &   & 1024 & \\
					\hline
					7 	&  F-TDNN   &t-5, t-2        &t-2, t+1   &t+1,t+4 &     &1024   &256 \\
					\hline
					8  &Dense  & & & & &1024 & \\
					\hline
					9 	&  F-TDNN    &t      &t       &t         & 5     &1024   &256 \\
					\hline
					10  &Dense  & & & & &1024 & \\
					\hline
					11 	&  F-TDNN    &t-5, t-2        &t-2, t+1   & t+1,t+4 &     &1024   &256 \\
					\hline
					12  &Dense  & & & & &1024 & \\
					\hline
					13 	&  F-TDNN   &t-5, t-2    &t-2,t+1     &t+1, t+4   & 3,7     &1024   &256 \\
					\hline
					14    &Dense  & & & & &1024 & \\
					\hline
					15 	&  F-TDNN    &t-5, t-2        &t-2, t+1   &t+1,t+4  &      &1024   &256 \\
					\hline
					16      &Dense  & & & & &1024 & \\
					\hline
					17	&  F-TDNN    &t  &t        &t     & 7,11,15    &1024    &256 \\
					\hline
					18 	&  Dense    &t      &    & &        &2048    & \\
					\hline
					19 	&  Dense    &t      &    & &        &2048    & \\
					\hline
					20 	&  Dense    &t      &    & &        &2048    & \\
					\hline
					21 	&  Pooling   &full-seq         & &  &  &  4096 & \\
					\hline
					22       &Dense    &  & &  &  &1024& \\
					\hline
					23     &Dense    &    & &     &     &1024 &\\
					\hline
					\multirow{2}{*}{24 }      &Dense-       & &    &  &  & N. spks.  &\\
					&Softmax      &    &  &  & &     &   \\
					\hline
				\end{tabular}
			\end{threeparttable}
		\end{center}
	\end{table*}

\subsection{resnet}
ResNet architecture is also based on tdnn x-vector \cite{Snyder2017Deep}. The five frame level tdnn layers in \cite{Snyder2017Deep} are replaced by ResNet34 (512 nodes) + DNN(512 nodes) + DNN(1000 nodes). Further details about ResNet34 can be found in \cite{ResNet16}. In our realization, acoustic features are regarded as a single channel picture and feed into the ResNet34. If the dimensions in the residual network don't match, zeros are added. The statistic pooling and segment level network stay the same. For the loss function, we use angular softmax with $m=4$. The x-vector is extracted from first DNN layer in segment level prior to the ReLU non-linearity. It has 512 dimensions.

	\subsection{multitask}
	\label{sec:multitask}
\begin{figure}[htb]
		\centering
		\centerline{\includegraphics[width=10cm]{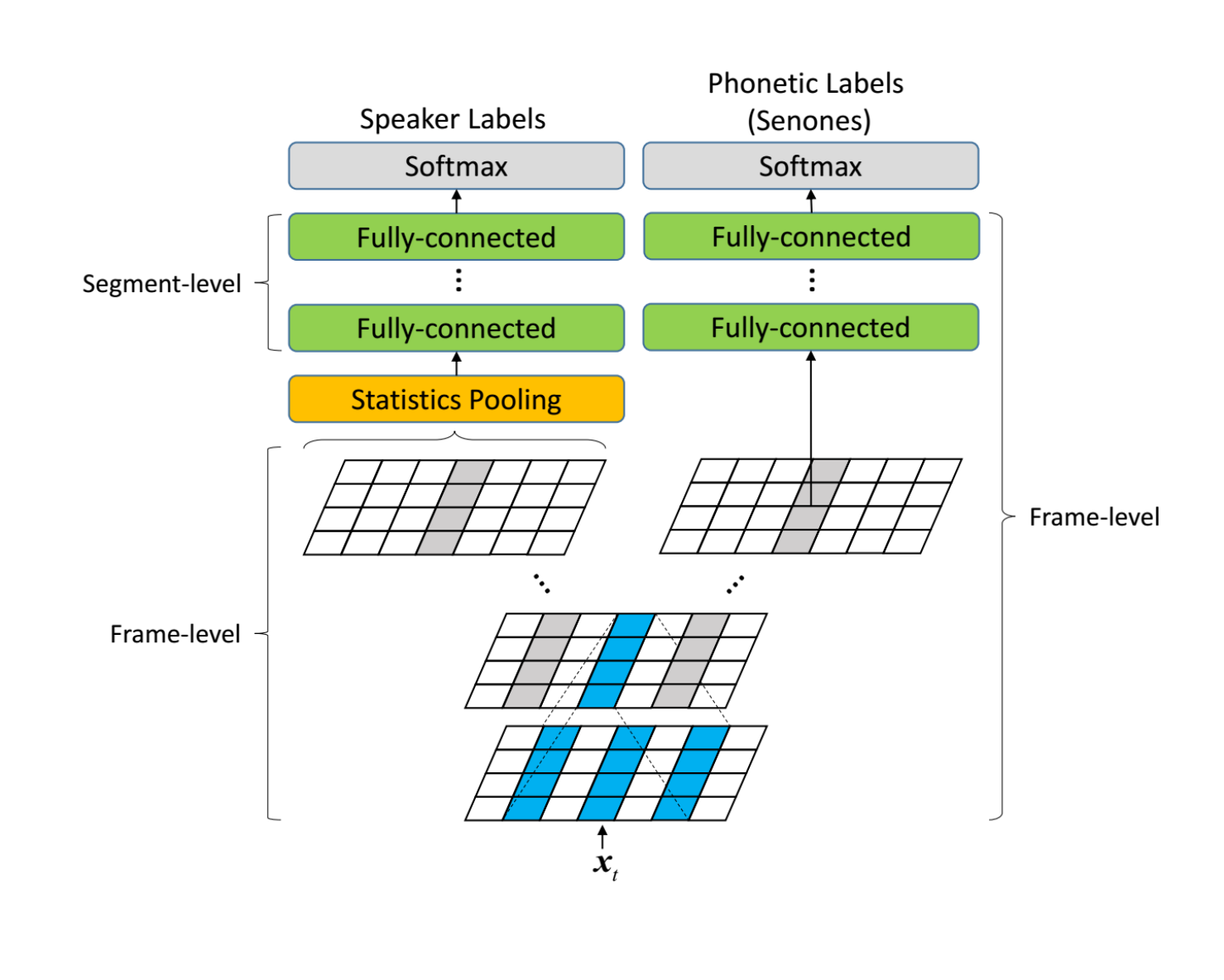}}
		\caption{multitask architecture for the speaker embedding extraction.}
		\label{fig:multitask}
	\end{figure}

	Multitask architecture is proposed in \cite{cvec2018}.
	It is a hybrid multi-task learning based on x-vector network and ASR network.
	It aims to introduce phonetic information by another neural acoustic model in ASR to help speaker recognition task.
	The architecture is shown in Fig. \ref{fig:multitask}.
	
	The frame-level part of the x-vector network is a 10-layer  TDNN.
	The input of each layer is the sliced output of the previous layer.
	The slicing parameter is: \{t - 2; t - 1; t; t + 1; t + 2\}, \{ t \}, \{ t - 2; t; t + 2 \}, \{t\}, \{ t - 3; t; t + 3 \}, \{t \}, \{t - 4; t; t + 4 \}, \{ t \}, \{ t \} , \{ t \}.
	It has 512 nodes in layer 1 to 9, and the 10-th layer has 1500 nodes.
	The segment-level part of x-vector network is a 2-layer fully-connected network with 512 nodes per layer.
	The output is predicted by softmax and the size is the same as the number of speakers.

	The ASR network has no statistics pooling component.
	The frame-level part of the x-vector network is a 7-layer TDNN.
	The input of each layer is the sliced output of the previous layer.
	The slicing parameter is: \{t - 2; t - 1; t; t + 1; t + 2\}, \{t - 2; t; t + 2\}, \{t - 3; t; t + 3\}, \{t\}, \{t\}, \{t\}, \{t\}.
	It has 512 nodes in layer 1 to 7.
	
	Only the first TDNN layer of the x-vector network is shared with the ASR network.
	The phonetic classification is done at the frame level, while the speaker labels are classified at the segment level.

	To train the multitask network, we need training data with speaker and ASR transcribed.
	But only Phonetic dataset fits this condition and the data amount is too small to train a neural network.
	So, we need to train a GMM-HMM speech recognition system to do phonetic alignment for other datasets.
	The GMM-HMM is trained using Phonetic dataset with features of 20-dimensional MFCCs with delta and delta-delta, totally 60-dimensional.
	The total number of senones is 3800.
	After training, forced alignment is applied to the SRE, Switchboard, and Voxceleb datasets using a fMLLR-SAT system.

	\subsection{c-vector}

	\begin{figure}[htbp]
		\centering
		\centerline{\includegraphics[width=10cm]{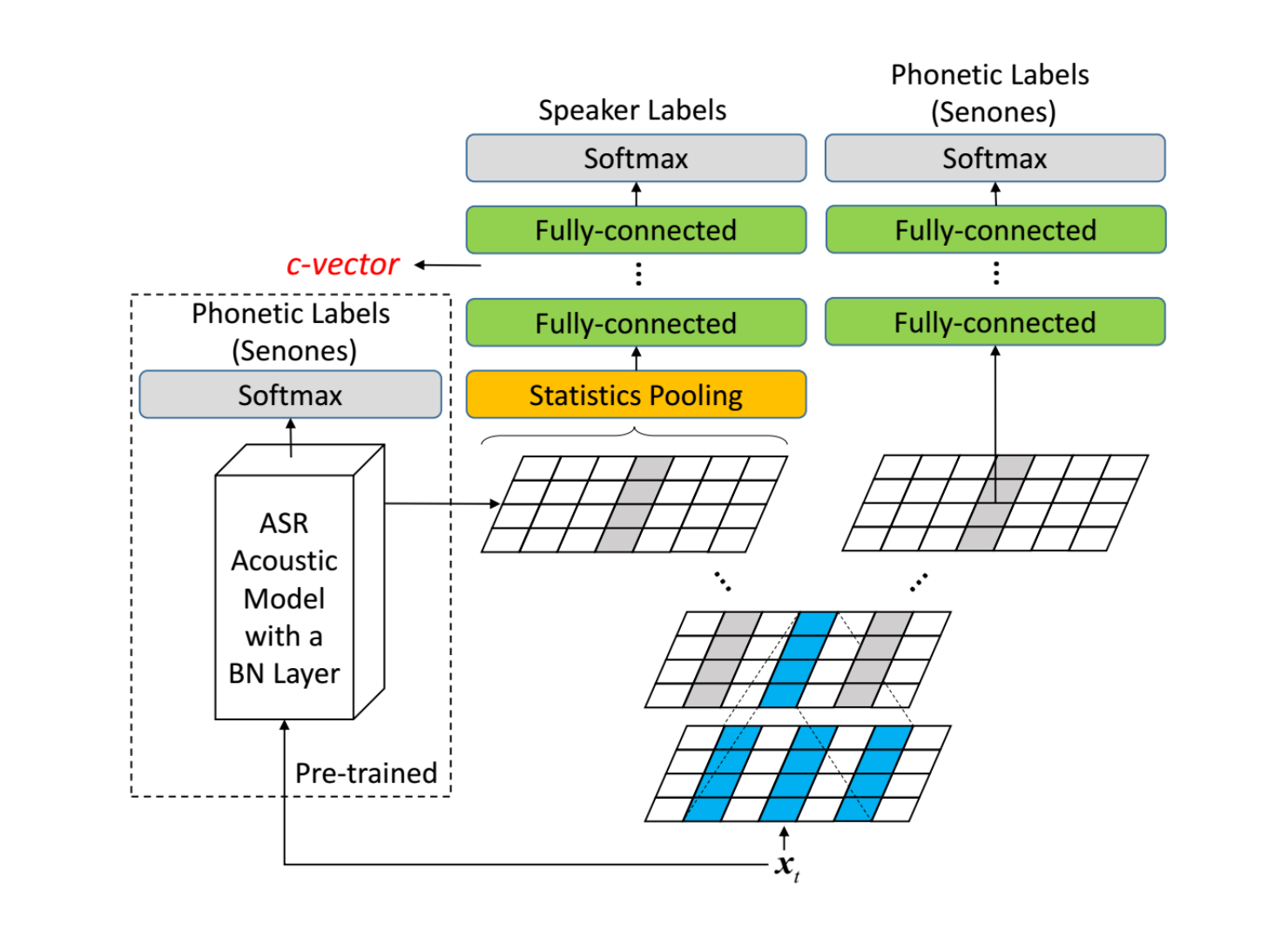}}
		\caption{multitask architecture for the speaker embedding extraction.}
		\label{fig:cvector}
	\end{figure}

	C-vector architecture is also one of our proposed systems in paper \cite{cvec2019}.
	As shown in Fig. \ref{fig:cvector}, it is an extension  of multitask architecture.
	It combines multitask architecture with an extra ASR Acoustic Model.
	The output of ASR Acoustic Model is concatenated with x-vector's frame-level output as the input of statistics pooling.
	Refer to \cite{cvec2019} for more details.

	The multitask part of c-vector has the same architecture as in the above section \ref{sec:multitask}
	ASR Acoustic Model of c-vector is a 5-layer TDNN network.
	The slicing parameter is  \{ t - 2; t - 1; t; t + 1; t + 2 \}, \{ t - 1; t; t + 1 \}, \{ t - 1; t; t + 1 \}, \{ t - 3; t; t + 3\}, \{ t - 6; t - 3; t\}.
	The 5-th layer is the BN layer containing 128 nodes and other layers have 650 nodes.
	
	A GMM-HMM is also trained as like in section \ref{sec:multitask} to do phonetic alignment for training datasets.
	
	


	\section{feature and back-end}
	
	23-dimensional MFCC (20-3700Hz)  is extracted as feature for etdnn/ams, ftdnn/as, eftdnn/ams, multitask and c-vector subsystems. 23-dimensional Fbank is used as feature for ResNet 16kHz subsystems. A simple energy-based VAD is used based on the C0 component of the MFCC feature \cite{kaldi2011}.
	
	For each neural network, its training data are augmented using the public accessible MUSAN and RIRS\_NOISES as the noise source. Two-fold data augmentation is applied for etdnn/ams, ftdnn/as, resnet, multitask and cvector subsystems. For eftdnn/ams subsystem, five-fold data augmentation is applied.

	After the embeddings are extracted, they are then transformed to 150 dimension using LDA.
	Then, embeddings are projected into unit sphere.
	At last, adapted PLDA with no dimension reduction is applied.
	
	The execution time is test on Intel Xeon E5-2680 v4.
	Extracting x-vector cost about 0.087RT.
	Single trial cost around 0.09RT.
	The memory cost about 1G for a x-vector extraction and a single trial.
	In the inference, only CPU is used.
	
	The speed test was performed on Intel Xeon E5-2680 v4 for etdnn\_ams, multitask, c-vector and ResNet system. Test on Intel Xeon Platinum 8168 for ftdnn and eftdnn system.
	Extracting embedding cost about 0.103RT for etdnn\_ams,  0.089RT for multitask, 0.092RT for c-vector, 0.132RT for eftdnn, 0.0639RT for ftdnn, and 0.112RT for ResNet.
	Single trial cost around 1.2ms for etdnn\_ams, 0.9ms for multitask, 0.9ms for c-vector, 0.059s for eftdnn, 0.0288s for ftdnn, 1.0ms for ResNet.
	The memory cost about 1G for an embedding extraction and a single trial.
	In the inference, we just use CPU.
	
	\section{Fusion}
	
	Our primary system is the linear fusion of all the above six subsystems by BOSARIS Toolkit on SRE19 dev and eval \cite{BOSARIS}.
	Before the fusion, each score is calibrated by PAV method (\em{pav\_calibrate\_scores}) on our development database.
	It is evaluated by the primary metric provided by NIST SRE 2019.

\begin{table}[t!]
	\centering
	\caption{Subsystem performance on SRE18 DEV and EVAL set. }
	\label{table:result}
	\resizebox{\columnwidth}{!}{
		\begin{tabular}{@{}ccccc@{}}
			\toprule
			\multirow{2}{*}{\textbf{System}} & \multicolumn{2}{c}{\textbf{SRE18 DEV}} & \multicolumn{2}{c}{\textbf{SRE18 EVAL}} \\ \cmidrule(l){2-5}
			        & \textbf{EER(\%)} & \textbf{min-DCF} & \textbf{EER(\%)} & \textbf{min-DCF} \\ \midrule
			etdnn & 3.95 & 0.222 & 2.59 & 0.198 \\
			ftdnn & 4.28 & 0.258 & 2.89 & 0.217 \\
			eftdnn & 3.67 & 0.196 & 2.56 & 0.204 \\
			resnet & 4.02 & 0.253 & 3.50 & 0.255 \\
			multitask & 4.35 & 0.276 & 3.58 & 0.278 \\
			c-vector & 3.92 & 0.252 & 3.10 & 0.249 \\ \midrule
			fused & 3.45 & 0.164 & 2.25 & 0.175 \\ \bottomrule
	\end{tabular}}
\end{table}	
	
	\bibliographystyle{IEEEbib}
	\bibliography{reference_for_sre19}
	
\end{document}